\documentstyle[twoside,fleqn,espcrc2,epsf]{article}

\newcommand{\be}{\begin{equation}}
\newcommand{\ee}{\end{equation}}
\newcommand{\order}{{\cal O}}

\newcommand{\eq}[1]{Eq.~(\ref{#1})}

\newcommand{\nl}{\nonumber \\}

\newcommand{\mev}{{\rm MeV}}
\newcommand{\gev}{{\rm GeV}}

\newcommand{\tslash}{{/ \hspace{-.5em} t}}
\newcommand{\uslash}{{/ \hspace{-.5em} U}}

\newcommand{\AmS}{{\protect\the\textfont2
A\kern-.1667em\lower.5ex\hbox{M}\kern-.125emS}}

\title{
Moving NRQCD
}

\author{
John~Sloan\address{Department of Physics and Astronomy, University of
Kentucky,
Lexington, KY 40506-0055, USA}${}^{,}$\thanks{.
This research was supported by DOE grants
DE-FG05-84ER40154 and DE-FC02-91ER75661 and by the Center for
Computational Sciences, University of Kentucky.}
}

\begin{document}

\begin{abstract}
I discuss the derivation and applications of Moving NRQCD (MNRQCD),
a generalization of NRQCD which allows the treatment of heavy
quarks moving with a finite velocity.  This formalism is vital in
reducing discretization errors in calculations of large recoil decays,
such as $B \rightarrow K^*\ + \ \gamma$ or the Isgur-Wise function
at large $w$.
\end{abstract}

\maketitle

\section{INTRODUCTION}

Many of the most interesting experimental probes of the Standard Model
involve exclusive weak decays of $B$ mesons.  Because the $B$ is much heavier 
than charmed and strange mesons, these decay products have large
recoil momenta.  This presents a problem for lattice 
calculations of these processes, since (naively) the spatial lattice
spacing must be fine enough to control discretization errors of order
$(ap_{recoil})$.  For example, in the process $B \rightarrow K^*\ + \ \gamma$,
the $K^*$ is kinematically constrained to have a momentum of about
$2.5\gev$ relative to the $B$; this means that an inverse lattice spacing
of $10\gev$ would be required just to reduce $(ap_{recoil})$ to 
about $25\%$.

A closely related problem has already been dealt with in lattice studies
of $B$ physics, namely that the {\it total } energy of the $b$ quark is
much larger than its (dynamically important) recoil momentum and kinetic
energy within the meson.  Several solutions have been found; the NRQCD and
FNAL approaches~\cite{cornell,IHW_refs}
are the most widely used.  In this talk I will discuss extending
the NRQCD method to heavy quarks moving with finite velocity; this
allows much of the recoil momentum to be removed from the calculation
before discretization.
The idea is that ``removing the rest mass'' in NRQCD 
is equivalent to shifting the 4-momentum of the $b$ quark, $P_b$ by an amount 
proportional to the time 4-vector, $\hat t$, while MNRQCD shifts the 
$b$ quark's 4-momentum by
a multiple of an arbitrary time-like 4-vector (preferably chosen to be 
the 4-velocity, $\hat U_B$, of the $B$ meson):
\begin{eqnarray}
{\rm  NRQCD} &:& P_b \rightarrow p_b + m_b\hat t,   \nl
{\rm MNRQCD} &:& P_b \rightarrow p_b + m_b\hat U_B,
\label{eq:mnrqcd_shift}
\end{eqnarray}
where $p_b$ is the shifted 4-momentum of the quark, which is relevant for 
lattice discretization errors, and $m_b$ is an arbitrarily chosen 
energy shift 
parameter (usually chosen non-relativistically so that the kinetic and 
static masses of the physical state are equal).\footnote{Note that 
$|\hat t |^2 = |\hat U_B |^2 = -1$ even when working with the Wick-rotated
Euclidean theory.} 
If $\hat U_B = \hat t$ the
two methods are identical, but when $\hat U_B \neq \hat t$ the MNRQCD
shift reduces the large spatial components of $P_b$.

Some work on this method has appeared previously.  In the same way
that MNRQCD is a generalization of NRQCD, the ``moving static'' formalism
of Mandula and Ogilvie~\cite{Mandula_Ogilvie} (MO) generalizes the static
theory, which in turn is the infinite mass limit of NRQCD.  Continuum 
HQET derivations are usually done in a Lorentz-covariant manner; the
terms in the continuum limit of the MNRQCD action at some order 
in $1/M$ should agree 
(up to field redefinitions) with the corresponding HQET action.
Finally, Hashimoto and 
Matsufuru~\cite{Hashimoto_Matsufuru} (HM) have derived the $\order(p^2/M)$
terms in the MNRQCD action and done numerical studies at low recoil;
Both MO and HM have also studied non-perturbative renormalization of
the input 4-velocity in the moving static and MNRQCD actions, respectively.
In this talk, I discuss the kinematics of specific high-recoil decays
and the expected improvement in discretization errors from using MNRQCD.
I sketch a method of deriving the tree-level MNRQCD
action by using the Foldy-Wouthuysen-Tani (FWT) transformation and
use this method to write down all $\order(1/M)$ terms in continuum MNRQCD.


\section{DECAY KINEMATICS}

The most illustrative decay to consider is $B \rightarrow K^*\ + \ \gamma$.
Because it is a two particle decay, the recoil 3-momentum of the $K^*$ (in
the rest frame of the $B$) is fixed to be $2.5\gev$, which corresponds
to an energy of $2.6\gev$ and a relativistic $\Gamma$ factor of $3.0$.  
The energy scale $\mu_{HAD}$ which controls discretization errors in Clover 
light hadron spectroscopy is a few hundred $\mev$; inverse lattice spacings
of $500$-$1000\mev$ give accuracies of a few 
percent~\cite{SCRI_coarse_lattice}.  In the $B$'s rest frame, however, 
$\mu_{K^*}$ will be controlled by $P_{K^*}$, the $K^*$'s 4-momentum.
This scale is about an order of magnitude larger, so inverse
lattice spacings of $5$-$10\gev$ might be required to achieve few percent
accuracies.

In contrast, consider the rest frame of the $K^*$. If we make this the
rest frame of our lattice discretization, then $\mu_{K^*}$ will be the
usual few hundred $\mev$, but the $B$ will be boosted to $\Gamma=3.0$;
$B$ meson discretization errors will dominate.  There are two sorts
of discretization errors we need to consider: 
``boosted muck''
and 
``moving mass''.
The boosted muck errors arise from Lorentz contraction; the brown muck
cloud around the $b$ quark will be Lorentz contracted in the direction of
motion when the $B$ meson is moving.  This should result in discretization
errors with a typical scale of $\Gamma_B\mu_{muck}$, where 
$\mu_{muck}\approx\mu_{HAD}$ is the $\mu$ for a $B$ meson at rest.

The moving mass errors are the same ones we encountered when 
boosting the $K^*$; a particle of mass $m$ has spatial momentum 
${\bf |p|} = m{\bf |v|}\Gamma = m\sqrt{\Gamma^2-1}$ when moving with velocity
${\bf v}$ and boost factor $\Gamma$.  After performing the MNRQCD
shift of the 4-momentum in \eq{eq:mnrqcd_shift}, the relevant mass
is $\bar\Lambda_B = m_B - m_b$, for which most NRQCD calculations
obtain a value of about $1\gev$.  The moving mass discretization scale in 
this case then is $\bar\Lambda_B\sqrt{\Gamma^2-1} \approx 3\gev$, i.e.\  
roughly the same as for the $K^*$ in the original frame.  This is
due to the numerical accident that $m_K^*\approx \bar\Lambda_B$; $v$ and 
$\Gamma$
don't care which meson is at rest so equal ``mass'' mesons have equal
momenta.

The problem in both the above frames is that the discretization scales
of the two mesons are very different.  An optimal frame is
one in which the two mesons are equally well discretized.
Since $m_K^*\approx \bar\Lambda_B$, this means that the two mesons 
should have roughly equal boost factors, i.e. 
$\Gamma_B = \Gamma_{K^*} = \sqrt{(1+\Gamma_{tot})/2} \approx \sqrt{2}$,
where $\Gamma_B$ and $\Gamma_{K^*}$ are the meson boosts relative to the
lattice frame and $\Gamma_{tot}$, their boost relative to each other,
has been take to be about $3$.  This reduction in $\Gamma$ by a factor
of $2$ leads to an even bigger improvement in the spatial momentum, which 
depends upon $\Gamma$ like $\sqrt{\Gamma^2-1}$.  The spatial momenta of 
the mesons in this frame are ${\bf p}_{K^*} \approx {\bf p}_B \approx 1\gev$, 
to be compared with the previous values of $2.5$-$3\gev$.  The boosted
muck errors in this frame are less than twice $\mu_{HAD}$; moving mass
errors should dominate.  Note that this is a worst-case analysis; the
moving mass errors of the $B$ meson might be further reduced
by using a different choice of $m_b$.

The other example I will consider is $B\rightarrow D^* + e + \nu$.  Since 
this is
a three particle decay, the $D^*$'s recoil momentum can range between $0$ and
$2.3\gev$.  The high recoil region gives much cleaner experimental
results for $e^+e^-$ colliders running at the $\Upsilon(4S)$, while
most of the theoretical work has been done near zero recoil.  If we
work in the rest frame of the $B$, then at maximum recoil the $D^*$
has 3-momentum, energy, and $\Gamma$ of $2.3\gev$, $3.0\gev$, and 1.1,
respectively.  If we use MNRQCD, however, we can subtract from the
4-momenta of {\it both} the $B$ and $D^*$ mesons.  Since $\bar\Lambda$
should be the same for the two mesons (i.e. about $1\gev$), again we
want to choose the frame where they have equal boost factors (which
will be about 1.025).  This implies a moving mass momentum of $220\mev$,
which is about the same size as the (not very) boosted muck scale.  This
means that, for a given lattice spacing, these runs should have the same
discretization errors as light spectroscopy calculations.  Even
if we decide not to use MNRQCD for the charm quark, the $D^*$'s spatial
momentum will still only be about $500\mev$.

\section{DERIVATION}

NRQCD is a (non-renormalizable) effective field theory which reproduces
the QCD heavy quark action.  As such, one should perform the normal
matching procedure to adjust the coefficients in the
action.  When one is working at tree-level, however, a (much simpler)
classical derivation can be used; this is just the FWT transformation.
The FWT transformation consists of going to a Dirac basis which diagonalizes
$\gamma_0$ and then, order by order in ${\bf p}/m$, 
block diagonalizing the fermion kernel into non-interacting quark
and anti-quark sectors.  This is equivalent to requiring that $\gamma_0$
commute with the fermion kernel.  The 4-momentum is then shifted
by rescaling the fermion fields by factors of $\exp{(m_b\gamma_0 t)}$.

To derive the tree-level MNRQCD action, one just needs to rewrite the
NRQCD derivation in covariant language.  This is done by writing the
time components of 4-vectors as their dot product with the unit time vector
(i.e.\ $\gamma_0$ is really $\hat \tslash$), while 3-vectors are obtained
with the ``spatial metric'' $s^{\mu\nu} = g^{\mu\nu} + t^\mu t^\nu$
(note that $h$ projects onto the spatial directions transverse to
$\hat t$, since $|\hat t|^2 = -1$).  The ${\bf p}^2/2m_0$ term in 
the NRQCD action, for example, can be written as 
$p^\mu s_{\mu\nu} p^\nu/2m_0$.  The MNRQCD derivation then consists of
going to a basis in which $\hat \uslash$, rather than $\hat \tslash$,
is diagonal and repeating the FWT transformation so that the fermion
kernel commutes with $\hat \uslash$.  The action obtained will be
the same as the NRQCD action (written in covariant form) with $\hat t$
replaced everywhere by $\hat U$.  Since the action kernel commutes
with $\hat \uslash$, the moving Pauli spinors can now be rescaled by
$\exp{(m_b\hat \uslash\hat U\cdot x)}$, which effects the subtraction
in \eq{eq:mnrqcd_shift}.  Dirac spinors (for use in currents) are recovered 
by performing the inverse fermion operations in reverse order: unrescaling
the fields, performing an inverse FWT transformation, and changing back
into a Dirac basis in which $\gamma_0$ is diagonal (this last is 
equivalent to boosting the fermion fields).

There are several subtleties in this procedure.  First of all, the
kinetic term, $p_\mu(\delta^{\mu\nu} + \hat U^\mu\hat U^\nu)p_\nu$,
contains both second temporal and spatio-temporal derivatives, which
could lead to spurious poles in the lattice dispersion relation.  The
solution is a field redefinition~\cite{cornell}; the equations of
motion are repeatedly used to eliminate time derivatives in higher-derivative
terms in favor of space derivatives.  This procedure has been carried
out in~\cite{Hashimoto_Matsufuru}; the result is quite nice:
\be
p_\mu(\delta^{\mu\nu} + \hat U^\mu\hat U^\nu)p_\nu
\rightarrow
p_i(\delta^{ij} - v^iv^j)p_j,
\ee
where $v^i = \hat U^i/\Gamma_{\hat U}$ is the velocity of the particle and
$i$, $j$ run over spatial directions.  This is a Lorentz contraction
term; when $|{\bf v}| = 1$ this term projects out the transverse momenta.

The other subtlety involves the other $\order(1/M)$ term in the NRQCD
action: $\sigma\cdot B/2m_0$.  The transcribed expression is in terms of
$\tilde\sigma_{\mu\nu}$, the commutator of the moving frame $\gamma$
matrices.  The problem is that we need to write this term using
$\sigma_{\mu\nu}$ (the commutator of the rest frame $\gamma$'s)
{\it without changing the spinor basis}.  This is achieved by boosting
only the vector indices.  One finds:
\begin{eqnarray}
&&\tilde\sigma\cdot\tilde B \rightarrow \\
&&\Gamma\left[{\bf \sigma}\cdot
            \left({\bf B} + {\bf E}\times{\bf v}\right)
  - {{\Gamma}\over{1+\Gamma}} \left({\bf \sigma}\cdot{\bf v}\right)
                            \left({\bf v}\cdot{\bf B}\right)
                            \right],\nonumber
\end{eqnarray}
where $\Gamma = \Gamma_{\hat U}$.  The first term is just the moving
$B$ field written in terms of rest frame $E$ and $B$, while the second
term again completes a transverse projector when $|v| = 1$ (the $E$ 
term is missing
because ${\bf v}\cdot{\bf E}\times{\bf v}$ vanishes).  The full
$\order(1/M)$ continuum MNRQCD action is then
\begin{eqnarray}
S      &=& \bar Q\left[ D_t + H_{MN} \right] Q \nl
H_{MN} &=& i{\bf v}\cdot{\bf D}
        -  {{D_i\left(\delta^{ij} - v^iv^j\right) D_j}\over{2m_0\Gamma}} \nl
       &+& {\sigma_i\left(\delta^{ij} - v^iv^j\right)
           \left({\bf B} + {\bf v}\times{\bf E}\right)_j}\over{2m_0}.
\end{eqnarray}


\end{document}